\documentclass[aps,prd,10pt,a4paper,superscriptaddress,preprintnumbers,showpacs,notitlepage,eqsecnum,nofootinbib]{revtex4-1}

\usepackage[english]{babel}
\usepackage{graphicx}
\usepackage{amsmath}
\usepackage{xcolor}
\usepackage{eufrak}
\usepackage{caption}
\usepackage{subcaption}
\def\sqr#1#2{{\vcenter{\vbox{\hrule height.#2pt\hbox{\vrule
width.#2pt height#1pt \kern#1pt\vrule width.#2pt}\hrule height.#2pt}}}}

\begin{document}

\title{Einstein-Gauss-Bonnet theory of gravity : \\
The Gauss-Bonnet-Katz boundary term\\}

\author{Nathalie Deruelle$^1$, Nelson Merino$^1$ and\ Rodrigo Olea$^2$\\
$^1$ \it{APC, CNRS-Universit\'e Paris 7,\\
75205 Paris CEDEX 13, France\\
$^2$Departamento de Ciencias F\'{\i}sicas, Universidad
Andres Bello,\\
Sazi\'e 2212, Piso 7, Santiago, Chile}\\
}

\begin{abstract}

We propose a boundary term to the Einstein-Gauss-Bonnet action for gravity, which uses the Chern-Weil theorem plus a dimensional continuation process, such that the extremization of the full action yields the equations of motion when Dirichlet boundary conditions are imposed. When translated into tensorial language, this boundary term is the generalization to this theory of the Katz boundary term and vector for general relativity. The boundary term constructed in this paper allows to deal with a general background and is not equivalent to the Gibbons-Hawking-Myers boundary term. However, we show that they coincide if one replaces the background of the Katz procedure by a product manifold. As a first application we show that this Einstein Gauss-Bonnet Katz action yields, without any extra ingredients, the expected mass of the Boulware-Deser black hole.

\end{abstract}

\maketitle

\section{Introduction}

The Gibbons-Hawking-York (GHY) boundary term \cite{GH,York} when added to the Einstein-Hilbert action for general relativity, and its generalization by Myers \cite{Myers} to the case of higher-dimensional Gauss-Bonnet and Lovelock theories of gravity, guarantees a \textit{Dirichlet variational principle}, that is that the extremization of the full action yields the Einstein / Gauss-Bonnet / Lovelock equations of motion when Dirichlet boundary conditions are imposed (see also \cite{GravanisWillison,Davis,DeruelleMadore}). 

As is well known, the Myers boundary terms can be written in the language of differential forms, making use of the Chern-Weil theorem \cite{ChernWeil} together with a \textit{dimensional continuation} procedure (see, e.g., Refs. \cite{Miskovic:2007mg,Mora:2006ka}).
The Chern-Weil theorem basically states that, given two gauge connection one-forms, the difference between two invariants constructed with their corresponding strength field is an exact form, i.e., the exterior derivative of an odd-form, which is called a \textit{transgression form}.
Transgression forms can be regarded as the generalization of Chern-Simons (CS) forms \cite{CS} by the inclusion of a second gauge field, and the Gibbons-Hawking-Myers (GHM) terms that define the Dirichlet problem in Lovelock gravity can be regarded as dimensional continuations of transgression forms for the Lorentz symmetry.
An essential feature of the procedure to write the GHM boundary terms in the vielbein formalism is that the second gauge field must be defined on a \textit{product manifold}, which is just an auxiliary manifold whose boundary has extrinsic curvature that is identically zero and coincides with the spacetime boundary. Thus, after using Gaussian coordinates, we recover the known expression which depends only on dynamical tensors of the boundary, and the Dirichlet problem is solved in a background-independent way.\footnote{Transgression forms have found physical applications in different contexts. For example, they were used in Ref. \cite{Anabalon:2006fj} to show that a four-dimensional gauged Wess-Zumino-Witten (WZW) Lagrangian arises from a five-dimensional Einstein-Gauss-Bonnet Lagrangian with special coefficients and that general relativity is a dynamical sector of this WZW theory where the symmetry is broken down to Lorentz. On the other hand, in Refs. \cite{TG1,TG2} it was proved that even-dimensional topological gravity \cite{Chamseddine} can be obtained from a CS and a transgression field theory invariant under the Poincaré group.}

A problem one has to deal with after having a well-posed variational principle with the GHM procedure is that the Dirichlet action and conserved charges obtained after applying Noether's theorem usually diverge. In the Einstein-Hilbert case the action can be regularized by the Hawking-Horowitz boundary term \cite{HH}, which just makes a background\footnote{By background we mean a vacuum solution (usually a maximally symmetric space) that is connected by a continuous parameter to the solution under study. We also remark that a product manifold cannot be a background (e.g., global AdS and Minkowski).} subtraction in the GHY term. When a negative cosmological constant is added, a background-independent regularization can be achieved by subtracting counterterms that depend on the intrinsic geometry of the boundary. This method, known as holographic renormalization, becomes technically involved in higher dimensions and a closed expression for these Dirichlet counterterms does not exist for a generic Lovelock gravity.
Remarkably, a universal regularization prescription for any Lovelock theory with anti–de Sitter (AdS) asymptotics was provided in Refs. \cite{Olea:2006vd,Kofinas:2007ns} using boundary terms that depend on the extrinsic curvature, also known as Kounterterms. This procedure modifies the boundary conditions, as it is the extrinsic curvature that is kept fixed at the boundary. However, as was shown in Ref. \cite{Miskovic:2006tm}, these kind of conditions arise naturally from the asymptotic form of the fields in the Fefferman-Graham expansion, and thus it is suitable to deal with the variational problem in a wide set of gravity theories that support asymptotically AdS solutions.

An alternative background-dependent boundary term to solve both the Dirichlet and regularization problems in Einstein-Hilbert theory was proposed by Katz \cite{Katz} (see also Ref. \cite{KBL}). There the Dirichlet  variational problem was solved by adding to the covariantized action the divergence of a vector which is constructed from the metric and the difference of the Christoffel symbols of the dynamical and background manifolds. 
The Katz boundary term ensures that the variational principle is well defined for Dirichlet boundary conditions and, together with a background subtraction, that Noether charges are finite. 
A nontrivial problem is finding a suitable Katz vector for each theory. For example, in Ref. \cite{DKO} a Katz-like vector for Einstein-Gauss-Bonnet (EGB) gravity was proposed in such a way that the expected mass as well as the expected thermodynamics of the Boulware-Deser black hole \cite{BoulwareDeser} and its rotating generalization (see Ref. \cite{DM}) were obtained. However, it was pointed out that the construction of the vector giving these results was not unique and that the proposed vector did not solve the Dirichlet problem.

In this work we follow a new route to propose alternative boundary terms, whose construction is also based on the structure of the Chern-Weil theorem.
These boundary terms still guarantee a Dirichlet variational principle and will allow us to write the Katz \cite{Katz} boundary term in the language of vielbeins, thus putting it on a footing similar to the Gibbons-Hawking-York boundary term.

We will then generalize the construction to the Gauss-Bonnet theory and it will be shown – in the language of differential forms – that this proposal solves the Dirichlet problem. When transposed into tensorial form, this will provide us with a unique generalization of the Katz vector for the Gauss-Bonnet action. As a first application, we will compare the obtained \textit{Einstein-Gauss-Bonnet-Katz} action to the one proposed in Ref. \cite{DKO} and show that it yields – without any extra ingredients – the expected mass and hence the expected thermodynamics of the Boulware-Deser black hole \cite{BoulwareDeser}.

This article is organized as follows. In Sec. II we review and compare different boundary terms  known in the literature, defining the Dirichlet problem in Einstein gravity. In Sec. III we show how these boundary terms can be written in the language of differential forms by introducing a special hybrid spin connection. We also show how these terms are related with the structure of the terms appearing in the two-dimensional Chern-Weil theorem. In Sec. IV A we review the Myers boundary term for Einstein-Gauss-Bonnet gravity and see that it can be regarded as the dimensional continuation of the transgression form appearing in the four-dimensional Chern-Weil theorem, when the second connection is associated with a product manifold. Then, in Sec. IV B we use the same theorem – but with the hybrid spin connection to show how the Katz-like vector must be defined. Finally, in Secs. IV C and IV D we show that our proposal solves the Dirichlet problem and that the associated Katz vector gives the right mass for the Boulware-Deser black hole.

\section{Einstein gravity with Dirichlet boundary terms, a recap \label{EinsteinBT_Dir}}

The Einstein-Hilbert action in $D$ dimensions is
\begin{equation}
I_{\rm E}={1\over16\pi}\int_{\cal M}\!\sqrt{-g}\,R\, d^Dx\label{EHaction}
\end{equation}
where Newton's constant and the speed of light are set equal to $1$, $g$ is the determinant of the components $g_{\mu\nu}$ of a pseudo-Riemannian metric in the coordinate system $x^{\mu}$ $=\{w,x^{i}\}$ (with $w$ being either the time or a \textquotedblleft radial" coordinate, and $x^{i}$ are coordinates in the boundary $\cal\partial M$ of dimension $d=D-1$), the integral is taken over some $D$-dimensional domain of integration ($d^{D}x=-\epsilon dwd^{d}x$, with $d^{d}x=dx^{i_{1}}\cdots dx^{i_{d}}$ and $\epsilon$ being $1$ or $-1$ when $w$
is, respectively, the \textquotedblleft radial" or time coordinate), and $R$ is the scalar curvature, $R=g^{\mu\nu}R_{\mu\nu}$, with $R_{\mu\nu}=R_{\ \mu\rho\nu}^{\rho}$ being the Ricci tensor, $R_{\ \nu\rho\sigma}^{\mu}=\partial_{\rho}\Gamma_{\nu\sigma}^{\mu}-\cdots$ being the Riemann tensor, and $\Gamma_{\nu\rho}^{\mu}$ are the Christoffel symbols.

The variation of $I_{\rm E}$ with respect to the contravariant components of the metric $\delta g^{\mu\nu}$ is\footnote{The calculation of $g^{\mu\nu}\delta R_{\mu\nu}$ is easily performed \textit{à la} Landau-Lifshitz in a locally inertial frame completed by covariantization to yield $\sqrt{-g}\,g^{\mu\nu}\delta R_{\mu\nu}=\partial_\mu\left[\sqrt{-g}\left(g^{\nu\rho}\delta\Gamma^\mu_{\nu\rho}-g^{\mu\nu}\delta\Gamma^\rho_{\nu\rho}\right)\right]$. The boundary term – $\sqrt{-g}\left(g^{\nu\rho}\delta\Gamma^w_{\nu\rho}-g^{w\nu}\delta\Gamma^\rho_{\nu\rho}\right)$ – is easily computed using Gaussian coordinates in which the metric reads $ds^2=\epsilon\, dw^2+h_{ij}\,dx^idx^j$ and  where the components of the extrinsic curvature are $K_{ij}={1\over2}\partial_wh_{ij}$.\\}
\begin{equation}
\delta I_{\rm E}={1\over16\pi}\int_{\cal M}\!\sqrt{-g}\,G_{\mu\nu}\,\delta g^{\mu\nu}\, d^Dx+{\epsilon\over16\pi}\int_{\cal\partial M}\sqrt{|h|}\left(K_{ij}-Kh_{ij}\right)\delta h^{ij}d^dx-{\epsilon\over8\pi}\int_{\cal\partial M}\delta\left(\sqrt{|h|}K\right)d^dx\,.
\label{EHactionVariation}
\end{equation}
Here $G_{\mu\nu}=R_{\mu\nu}-{1\over2}g_{\mu\nu}R$ is the Einstein tensor ; $\epsilon=-1$ if the boundary $\cal\partial M$ of $\cal M$ is spacelike and $\epsilon=1$ if it is timelike, $h$ is the determinant of the induced metric on $\cal\partial M$ with components $h_{ij}$, and $K_{ij}$ is the extrinsic curvature of $\cal\partial M$ and $K=h^{ij}K_{ij}$ is its trace.\\

In order to build a Dirichlet variational principle – that is, in order for the extremization of the action for gravity to yield the Einstein (vacuum) equations of motion $G_{\mu\nu}=0$ when Dirichlet boundary conditions are imposed, i.e., when $\delta h^{ij}=0$ on $\partial\cal M$ – appropriate boundary terms must be added to the Einstein-Hilbert action. The two examples we will concentrate upon are~: \\

$\bullet$ The Einstein Gibbons-Hawking-York (EGHY) action \cite{GH,York}
\begin{equation}
I_{\rm EGHY}=I_{\rm E}+I_{\rm GHY}\quad\hbox{with}\quad I_{\rm GHY}={\epsilon\over8\pi}\int_{\cal\partial M}\!\sqrt{|h|}Kd^dx
\label{EGHYaction}
\end{equation}
which yields a  Dirichlet variational principle since, on shell (that is when $G_{\mu\nu}=0$)~:
\begin{equation}
\delta I_{\rm EGHY}\vert_{\rm on shell}={\epsilon\over16\pi}\int_{\cal\partial M}\!\sqrt{|h|}(K_{ij}- Kh_{ij})\delta h^{ij}d^dx
\label{EGHYactionVariation}
\end{equation}
which vanishes if Dirichlet boundary conditions ($\delta h^{ij}=0$ on $\partial\cal M$) are imposed. 
As mentioned in the Introduction, the finite action is obtained after making a background subtraction in the GHY term. Thus, one obtains the Einstein Hawking-Horowitz (EHH) action \cite{HH},
\begin{equation}I_{\rm EHH}=I_{\rm E}+I_{\rm HH}\quad\hbox{with}\quad I_{\rm HH}={\epsilon\over8\pi}\int_{\cal\partial M}\!\sqrt{|h|}(K-\bar K)d^dx\,\label{EHHaction}
\end{equation}
where $\bar K=\bar h^{ij}\bar K_{ij}$ is the trace of the extrinsic curvature of a background manifold with metric $\bar h_{ij}$ on its boundary and extrinsic curvature $\bar K_{ij}$. It also yields a Dirichlet variational principle since, on shell~:
\begin{equation}
\delta I_{\rm EHH}\vert_{\rm on shell}={\epsilon\over16\pi}\int_{\cal\partial M}\!\sqrt{|h|}\left[K_{ij}- (K-\bar K)h_{ij}\right]\delta h^{ij}d^dx\,.
\label{EHHactionVariation}
\end{equation}

$\bullet$ The Einstein Katz (EK) action \cite{Katz,KBL}
\begin{equation}
I_{\mathrm{EK}}=I_{\mathrm{E}}+I_{\mathrm{E}}^{\mathrm{K}}-\bar{I}%
_{\mathrm{E}}\quad\hbox{with}\quad I_{\mathrm{E}}^{\mathrm{K}}={\frac{1}%
{16\pi}}\int_{\mathcal{M}}\!\partial_{\mu}\left(  \sqrt{-g}\,k_{\mathrm{E}%
}^{\mu}\right)  d^{D}x\quad\hbox{and}\quad\bar{I}_{\mathrm{E}}={\frac{1}%
{16\pi}}\int_{\bar{\mathcal{M}}}\!\sqrt{-\bar{g}}\,\bar{R}\,d^{D}%
x\,,\label{EKaction}%
\end{equation}
where the vector $k_{\rm E}^\mu$ is defined as \cite{Katz}
\begin{equation}
k_{\rm E}^\mu=-(g^{\nu\rho}\Delta^\mu_{\nu\rho}-g^{\mu\nu}\Delta^\rho_{\nu\rho})\quad\hbox{with}\quad \Delta^\mu_{\nu\rho}=\Gamma^\mu_{\nu\rho}-\bar\Gamma^\mu_{\nu\rho}\,,\label{Kvector}
\end{equation}
and where $\bar\Gamma^\mu_{\nu\rho}$ are the Christoffel symbols of the background metric $\bar g_{\mu\nu}$. 
To understand the origin of this vector, we recall that the EH Lagrangian density can be written as $\sqrt{-g}R=\sqrt{-g}G+\partial _{\mu }\left( \sqrt{-g}v^{\mu }\right) $,
where $G=g^{\mu \nu }\left( \Gamma _{\mu \rho }^{\lambda }\Gamma _{\nu \lambda }^{\rho }-\Gamma _{\mu \nu }^{\rho }\Gamma _{\rho \lambda }^{\lambda }\right) $ and
$v^{\mu }=g^{\nu \rho }\Gamma _{\nu \rho }^{\mu }-g^{\mu \nu }\Gamma _{\nu \rho }^{\rho }$. Then, a variational principle with no boundary term would be obtained by just subtracting the divergence $\partial _{\mu }\left( \sqrt{-g}v^{\mu }\right) $. However, this leads to an action that is not invariant under diffeomorphisms. Using $\delta \bar g_{\mu \nu}=0$ and that the difference of two Christoffel symbols is a tensor, Katz constructed the vector $k_{\rm E}^\mu$ as a covariantized version of the vector $v^{\mu }$ and showed that adding its divergence to the EH action allows to obtain a well-posed variational principle for Dirichlet boundary conditions.
Indeed, using Gauss coordinates the Katz boundary term reads
\begin{equation}
I^{\rm K}_{\rm E}={1\over16\pi}\int_{\partial\cal M}\!\sqrt{|h|}\,k^w_{\rm E}\,d^dx={\epsilon\over8\pi}\int_{\partial\cal M}\sqrt{|h|}\!\left[(K-\bar K)-{1\over2}(h^{ij}-\bar h^{ij})\bar K_{ij}\right]d^dx\,.\label{KboundaryTerm}
\end{equation}
and thus, the on-shell variation of the Einstein Katz action is
\begin{equation}
\delta I_{\rm EK}\vert_{\rm on shell}={\epsilon\over16\pi}\int_{\cal\partial M}\!\sqrt{|h|}\left[(K_{ij}-\bar K_{ij})- h_{ij}(K-\bar K)+{h_{ij}\over2}(h^{kl}-\bar h^{kl})\bar K_{kl}\right]\delta h^{ij}d^dx
\label{EKactionVariation}
\end{equation}
which clearly vanishes if Dirichlet boundary conditions are imposed.\\

One notes that the boundary terms to define the Dirichlet actions $I_{\rm EGHY}$ and $I_{\rm EK}$ are not equivalent. The first difference is that the first one is background independent while the second one is not. On the other hand, the GHY boundary term $I_{\rm GHY}$ is covariant only with respect to the boundary and as a consequence its variation cancels only the variation of the metric derivatives which are normal to the boundary. Instead, the Katz boundary term $I^{\rm K}_{\rm E}$ can be written as a fully covariant expression (\ref{EKaction}) and is such that its variation cancels the variation of all of the metric derivatives. Besides, as was shown in Ref. \cite{K_null}, the Katz procedure is also useful for applying the variational principle when boundaries are null. However, this is not the problem we treat in this paper.

Finally, we note that $I_{\rm EK}$ reduces to the action $I_{\rm EGHY}$ if the background is taken to be a product manifold. Indeed, in that case the line element in Gaussian coordinates is $ds^2=\epsilon dw^2+\bar h_{ij}(x^l)\,dx^i\,dx^j$, and thus $\bar K_{ij}=0$ because the metric $h_{ij}$ on the boundary does not depend on $w$.

\section{Einstein gravity and boundary terms in the vielbein formalism}
\label{sectionIII}

Another way to address the variational problem in Einstein's gravity is to use the vielbein formalism. Here one switches from the previous coordinate one-form basis of the
cotangent spaces ($\mathrm{d}x^{\mu}$, with $\mathrm{d}$ being the exterior
derivative acting on a function $f$ as $\mathrm{d}f=\partial_{\alpha
}f\mathrm{d}x^{\alpha}$ and such that $\mathrm{d}\mathrm{d}x^{\rho}=0$) to a
tetrad one-form basis ($e^{A}$) such that the metric tensor $\mathrm{g}%
=g_{\mu\nu}\mathrm{d}x^{\mu}\otimes\mathrm{d}x^{\nu}$ (where $\otimes$ is the
tensorial product operator and where $\mathrm{g}(\partial_{\mu},\partial_{\nu
})=g_{\mu\nu}$, with $\partial_{\mu}$ being the conjugate coordinate basis of the
tangent spaces) is diagonalized into $\mathrm{g}=\eta_{AB}e^{A}\otimes e^{B}$,
where $\eta_{AB}$ is the Minkowski metric. Hence, $A,B=0,1,\ldots$ are Lorentz
indices which are moved with $\eta_{AB}$ and its inverse, $\omega_{\ B}^{A}$
is the (torsionless) spin connection, defined by $\mathrm{d}e^{A}+\omega_{\ B}^{A}\wedge e^{B}=0$
where a wedge denotes the exterior product (that is, the antisymmetrized tensorial product),
$\varepsilon_{AB}$ is the (Lorentz) Levi-Civita index such that $\varepsilon_{01}=1$, and finally $\Omega^{AB}=\eta^{BC}\Omega_{\ C}^{A}$, where
$\Omega_{\ B}^{A}=\mathrm{d}\omega_{\ B}^{A}+\omega_{\ C}^{A}\wedge
\omega_{\ B}^{C}$ is the curvature two-form. (And similar definitions for the
\textit{barred} background manifold.)\\

A convenient starting point is the Chern-Weil (CW) theorem in two dimensions (see, e.g., Ref. \cite{ChernWeil}), which states that
\begin{equation}
\varepsilon_{AB}(\Omega^{AB}-\bar{\Omega}^{AB})=d(  \varepsilon_{AB}\,\bar{\theta}^{AB})\quad\hbox{where}\quad \bar\theta^{AB}=
\omega^{AB}-\bar\omega^{AB}\,,\label{ChernWeilTheorem2}
\end{equation}
where $\omega_{\ B}^{A}$ and $\bar{\omega}_{\ B}^{A}$ are two given spin connections with curvature two-forms $\Omega_{\ B}^{A}$ and $\bar{\Omega}_{\ B}^{A}$, $\varepsilon_{AB}$ is the Lorentz Levi-Civita tensor, and the term inside the exterior derivative – $\varepsilon_{AB}\,\bar{\theta}^{AB}$ – is called the transgression form.
This theorem (which can be seen as an identity in two dimensions) suggests 
introducing as a four-dimensional \textit{Einstein-Chern-Weil} (ECW) action its dimensional continuation,  $I_{\rm ECW}$, defined as
\begin{equation}
I_{\mathrm{ECW}}=\frac{1}{32\pi}\int_{\mathcal{M}_{4}}\varepsilon_{ABCD}\,\Omega^{AB}\wedge e^{C}\wedge e^{D}-\frac{1}{32\pi}\int_{{\mathcal{M}}_{4}}{\rm d}\left(\varepsilon_{ABCD}\,\bar{\theta}^{AB}\wedge e^{C}\wedge e^{D}\right)  \,. 
\label{ChernWeilAction4}
\end{equation}
where the first term yields the Einstein-Hilbert action (\ref{EHaction}). The translation uses the following relations: 
\begin{equation}
\begin{aligned}
\Omega^{AB}=\frac{1}{2}e_{\alpha}^{A}e^{B\beta}R_{\ \beta\mu\nu}^{\alpha}\mathrm{d}x^{\mu}\wedge\mathrm{d}x^{\nu}\,, \ \ \ \varepsilon_{ABCD}e_{\mu}^{A}e_{\nu}^{B}e_{\rho}^{C}e_{\sigma}^{D}=\sqrt{-g}\varepsilon_{\mu\nu\rho\sigma}
\cr
\mathrm{d}x^{\beta}\wedge\mathrm{d}x^{\gamma}\wedge\mathrm{d}x^{\rho}\wedge\mathrm{d}x^{\sigma}=-\varepsilon^{\beta\gamma\rho\sigma}d^{4}x \,, \ \  \varepsilon_{\mu\nu\rho\sigma}\varepsilon^{\beta\gamma\rho\sigma}=-2(\delta_{\mu}^{\beta}\delta_{\nu}^{\gamma}-\delta_{\nu}^{\beta}\delta_{\mu}^{\gamma})\,
\label{id_translation}
\end{aligned}
\end{equation}
where $x^{\mu}=t,r,\phi,\varphi \,$, $\delta_{\beta}^{\alpha}$ is the Kronecker symbol and $\varepsilon_{\mu\nu\rho\sigma}$ is the Levi-Civita symbol with spacetime indices such that $\varepsilon_{tr\phi\varphi}=1$, $\varepsilon^{tr\phi\varphi}=-1$.
Besides, the possible background bulk term $\int\varepsilon_{ABCD}\,\bar\Omega^{AB}\wedge e^C\wedge e^D$ must be omitted in order to not spoil the field equations.\\

Now, as is well known (see, e.g., Ref. \cite{Myers}) Eq. (\ref{ChernWeilAction4}) reproduces the Einstein Gibbons-Hawking-York action (\ref{EGHYaction}) if the spin connection $\bar\omega^{AB}$ is associated with a product manifold whose line element in Gaussian coordinates is $d\bar{s}^{2}=\epsilon\,dw^{2}+\bar{h}_{ij}(x^{l})dx^{i}dx^{j}$, where the metric $\bar{h}_{ij}$ on the boundary does not depend on the coordinate normal to the boundary $w$. Then, the extrinsic curvature of the background boundary vanishes, $\bar{K}_{ij}=0$, and all of the terms containing barred quantities disappear. In this case, the translation of Eq. (\ref{ChernWeilAction4}) to tensorial language can only be performed using Gaussian coordinates, i.e., if the covariance is reduced to the boundary.
Indeed, in Gaussian coordinates the Lorentz indices are split as $A=\left(\mathrm{n},a\right)  $ (where $\mathrm{n}$ is, respectively, 0 or 1 when $\epsilon$ is $1$ or $-1$), and for $\bar{\omega}$ associated with a product manifold we have the following properties:
\begin{equation}
\begin{aligned}
\bar{\theta}^{ab}=\omega^{ab}-\bar{\omega}^{ab}=0\,, \quad \omega^{\mathrm{n}a}=-\omega^{a\mathrm{n}}=-\epsilon K^{a} \,,
\cr 
\text{with} \quad K^{a}=e_{i}^{a}K_{j}^{i}dx^{j} \quad \text{and}\quad  \bar{\omega}^{\mathrm{n}a}=-\bar{\omega}^{a\mathrm{n}}=-\epsilon\,\bar{e}_{i}^{a}\bar{K}_{j}^{i}dx^{j}=0\,. 
\label{id_translation_GHY}
\end{aligned}
\end{equation}
Then, using Gauss' theorem for a three-form $Q$ as $\int_{\mathcal{M}}dQ=-\epsilon\int_{\partial\mathcal{M}}Q\,$, the boundary term in Eq. (\ref{ChernWeilAction4}) reads
\begin{equation}
-\int d\left(  \varepsilon
_{ABCD}\,\bar{\theta}^{AB}\wedge e^{C}\wedge e^{D}\right)  =\epsilon
\int2\varepsilon_{abc}\,K^{a}\wedge e^{b}\wedge e^{c}=\epsilon\int4\sqrt
{h}Kd^{3}x\,, 
\end{equation}
where the last equality is obtained using
\begin{equation}
\varepsilon_{\mathrm{n}abc}=-\epsilon\,\varepsilon_{abc}\,, \quad \varepsilon_{abc}e_{i}^{a}e_{j}^{b}e_{k}^{c}=\sqrt{\left\vert h\right\vert}\varepsilon_{ijk}\,, \quad dx^{i}dx^{j}dx^{k}=-\epsilon\,\varepsilon^{ijk}d^{3}x\,, \quad  \varepsilon_{ljk}\,\varepsilon^{ijk}=-2\epsilon\delta_{j}^{l}\,.
\label{id_translation_GHY_2}
\end{equation}
\smallskip

On the other hand, a fully covariant translation to tensorial language of the boundary term in Eq. (\ref{ChernWeilAction4}) is not possible in general. 
The reason is that after using the \textit{tetrad postulate}, namely, $\omega_{\ B}^{A}=\left(  e_{\alpha}^{A}e_{B}^{\gamma}\Gamma_{\nu\gamma}^{\alpha}+e_{\alpha}^{A}\partial_{\nu}e_{B}^{\alpha}\right)  dx^{\nu}$ (and a similar expression with bars for the second connection), there is no general way to get rid of the vielbeins to obtain an expression depending on the metrics and Christoffel symbols only.
However, there is a case that is general enough for which a covariant translation is possible. To show this, we first introduce the \textit{hybrid} spin connection as
\footnote{The geometric properties and the proof that $\tilde{\omega}^{AB}$ does transform as a spin connection under local Lorentz transformations can be found in Ref. \cite{MerinoOlea}.\\}
\begin{equation}
\tilde{\omega}_{\ B}^{A}\equiv\left(  e_{\alpha}^{A}e_{B}^{\beta}\bar{\Gamma}_{\nu\beta}^{\alpha}+e_{\alpha}^{A}\partial_{\nu}e^{B\alpha}\right)  dx^{\nu},
\label{hybrid_lorentz}
\end{equation}
where $\bar{\Gamma}$ is the Christoffel symbol of an \textit{a priori} arbitrary background manifold $\bar {\cal M}$. 
Then, the difference of the connections $\omega^{AB}=\eta^{BC}\omega_{\ C}^{A}$ and $\tilde{\omega}^{AB}=\eta^{BC}\tilde{\omega}_{\ C}^{A}$ is given by 
\begin{equation}
\tilde{\theta}^{AB}=\omega^{AB}-\tilde{\omega}^{AB}=e_{\mu}^{A}e_{\nu}^{B}\tilde{\theta}^{\mu\nu}
\label{theta_hyb_def}
\end{equation}
where $\tilde{\theta}^{\mu\nu}=\omega^{\mu\nu}-\tilde{\omega}^{\mu\nu}$ with
\begin{equation}
\omega^{\mu\nu}\equiv g^{\nu\alpha}\omega_{\ \alpha}^{\mu
}\quad,\quad\omega_{\ \alpha}^{\mu}=\Gamma_{\beta\alpha}^{\mu}\,\mathrm{d}
x^{\beta}\quad
\text{and}\quad\tilde{\omega}^{\mu\nu}\equiv g^{\nu\alpha}
\bar{\omega}_{\ \alpha}^{\mu}\quad,\quad\bar{\omega}_{\ \alpha}^{\mu}
=\bar{\Gamma}_{\beta\alpha}^{\mu}\,\mathrm{d}x^{\beta}\,.
\label{other_hyb_def}
\end{equation}
If we use the hybrid connection (\ref{hybrid_lorentz}) in the ECW action (\ref{ChernWeilAction4}), then this action can be translated (using the properties given in Eq. (\ref{id_translation})) into a fully covariant coordinate basis, namely,
\begin{align}
\tilde{I}_{\mathrm{ECW}}  & =\frac{1}{32\pi}\int_{\mathcal{M}_{4}}%
\varepsilon_{ABCD}\,\Omega^{AB}\wedge e^{C}\wedge e^{D}-\frac{1}{32\pi}%
\int_{{\mathcal{M}}_{4}}\mathrm{d}\left(  \varepsilon_{ABCD}\,\tilde{\theta
}^{AB}\wedge e^{C}\wedge e^{D}\right)  
\label{tildeECWactionForm}\\
& ={\frac{1}{32\pi}}\int_{\mathcal{M}_{4}}\sqrt{-g}\,\varepsilon_{\mu\nu
\rho\sigma}\,\Omega^{\mu\nu}\wedge\mathrm{d}x^{\rho}\wedge\mathrm{d}x^{\sigma
}-{\frac{1}{32\pi}}\int_{{\mathcal{M}_{4}}}\mathrm{d}\left(  \sqrt{-g}\,\varepsilon_{\mu\nu\rho\sigma}\, \tilde{\theta}^{\mu\nu}\wedge\mathrm{d}x^{\rho}\wedge
\mathrm{d}x^{\sigma}\right)  
\label{tildeECWactionForm_ten}%
\end{align}
with $\Omega^{\mu\nu}=g^{\nu\alpha}\Omega_{\ \alpha}^{\mu}$ and $\Omega_{\ \alpha}^{\mu}={\frac{1}{2}}R_{\ \alpha\beta\gamma}^{\mu}\mathrm{d}x^{\beta}\wedge\mathrm{d}x^{\gamma}\,$.
Hence the use of the hybrid connection (\ref{hybrid_lorentz}) provides a way to translate the boundary term in the ECW action (\ref{tildeECWactionForm}) into a fully covariant coordinate basis (\ref{tildeECWactionForm_ten}).\\

These remarks pave the way to write the Einstein-Katz action in the vielbein language.
First, we notice that the definitions given in Eqs. (\ref{theta_hyb_def}) and (\ref{other_hyb_def}) lead to $\tilde{\theta}^{AB}=e_{\mu}^{A}e_{\nu}^{B}g^{\nu\alpha}\Delta_{\beta\alpha}^{\mu}\mathrm{d}x^{\beta}$, where $\Delta_{\beta\alpha}^{\mu}=\Gamma_{\beta\alpha}^{\mu}-\bar{\Gamma}_{\beta\alpha}^{\mu}$ is the tensor appearing in the definition of the Katz vector (\ref{Kvector}).
Hence, the Katz vector is related to the exterior derivative of the boundary term in Eq. (\ref{tildeECWactionForm}) 
as\footnote{Consistency between the fully covariant versions of the Katz action (\ref{EKaction}) and (\ref{tildeECWactionForm})-(\ref{dbetaEqualDivKatz}) and its expression in Gaussian coordinates (\ref{KboundaryTerm}) is obtained by taking into account that, with our conventions, the Gauss' theorems for a vector $A^{\mu}$ and a three-form $Q$ are given, respectively, by $\int_{\mathcal{M}_{4}}\partial_{\mu}\left(  \sqrt{-g}A^{\mu}\right)  d^{4}x=\epsilon\int_{\partial\mathcal{M}_{4}}\sqrt{\left\vert h\right\vert }n_{\mu}A^{\mu}d^{4}x$ and $\int_{\mathcal{M}_{4}}dQ=-\epsilon\int_{\partial\mathcal{M}_{4}}Q$.\\
}
\begin{equation}
-{\rm d}(\sqrt{-g}\,\varepsilon_{\mu\nu\rho\sigma}\,\tilde\theta^{\mu\nu}\wedge {\rm d}x^\rho\wedge {\rm d}x^\sigma)=2\partial_\mu(\sqrt{-g}\, k_{\rm E}^\mu)\,d^4x\,. 
\label{dbetaEqualDivKatz}
\end{equation}
Therefore, the ECW action (\ref{tildeECWactionForm}) reproduces the Einstein Katz action (\ref{EKaction}), namely, $I_{\rm EK}=\tilde I_{\rm ECW}\,$.

Interestingly, if $\bar{\Gamma}$ is associated with a product manifold, then the relation (\ref{dbetaEqualDivKatz}) still holds and the action (\ref{tildeECWactionForm}) represents a fully covariant way to write the EGHY action (\ref{EGHYaction}).
Indeed, showing the equality of the boundary terms in Eqs. (\ref{tildeECWactionForm}) and (\ref{EGHYaction}), that is, that 
\begin{equation}
\sqrt{-g}\,\varepsilon_{\mu\nu\rho\sigma}\,\tilde{\theta}^{\mu\nu}\wedge\mathrm{d}x^{\rho}\wedge\mathrm{d}x^{\sigma}=4\sqrt{|h|}Kd^{3}x \,,
\label{Cov_GHY}
\end{equation}
is easily done using Gaussian coordinates: $ds^{2}=\epsilon\,dw^{2}+h_{ij}dx^{i}dx^{j}$ and $d\bar{s}^{2}=\epsilon\,dw^{2}+\bar{h}_{ij}(x^{l})dx^{i}dx^{j}$. In these coordinates $\Gamma_{ij}^{w}=-\epsilon
K_{ij}$ and $\Gamma_{iw}^{j}=K_{i}^{j}$ where $K_{ij}={\frac{1}{2}}%
\partial_{w}h_{ij}$ is the extrinsic curvature of $\partial\mathcal{M}$, $K_{i}^{l}=h^{lj}K_{ij}$ and $\bar{K}_{ij}=0$. Hence the indices $\rho$ and $\sigma$ on the lhs of Eq. (\ref{Cov_GHY}) reduce to $j$ and $k$. We also used the relations $\varepsilon_{wljk}=-\varepsilon
_{lwjk}=-\epsilon\,\varepsilon_{ljk}$ and $\mathrm{d}x^{i}\wedge\mathrm{d}x^{j}\wedge\mathrm{d}x^{k}=-\epsilon\,e^{ijk}d^{3}x$, as well as
$\varepsilon_{ljk}\varepsilon^{ijk}=-2\epsilon\delta_{l}^{i}$. 

The fact that Eq. (\ref{tildeECWactionForm}) also represents a fully covariant way to write the EGHY action if the background is taken to be a product manifold is completely consistent with the fact that the Katz action (\ref{KboundaryTerm}) leads to the GHY action (\ref{EGHYaction}) under the same condition.
Thus, the use of the hybrid connection (\ref{hybrid_lorentz}) allows us to construct $\tilde I_{\rm ECW}$, which reproduces: a) the EK action when $\bar{\Gamma}$ is associated with an arbitrary background, and b) the EGHY action when $\bar{\Gamma}$ is associated with a product manifold.

The expressions (\ref{tildeECWactionForm})-(\ref{Cov_GHY}) are the first results of this paper, which bridge the gap between the Gibbons-Hawking-York and Katz boundary terms.
Additionally, one can easily check that the Hawking-Horowitz (HH) action (\ref{EHHaction}) can be written using differential forms as Eq. (\ref{tildeECWactionForm_ten}) if one replaces $\tilde\omega^{\mu\nu}$ by $\bar\omega^{\mu\nu} = \bar g^{\nu\alpha}\bar\omega^\mu_{\ \alpha}\,$; but then, that action cannot be written in the vielbein language as Eq. (\ref{tildeECWactionForm}), and thus the HH action does not correspond to a translation of the ECW action for any given pair of spin connections. This shows that the HH action is not related to a fundamental geometrical object, such as a transgression (as defined in Eq. (\ref{ChernWeilTheorem2})). 
This is probably why no generalization of the HH action is known so far in the literature for the Gauss-Bonnet and Lovelock cases.

 \section{The Gauss-Bonnet-Katz action and vector}

 \subsection{The Gauss-Bonnet-Myers action: A recap}

 $\bullet$ {\it In tensorial language}
 \medskip

 The Gauss-Bonnet action is \cite{Lanczos,Lovelock,DeruelleMadore}
\begin{equation}
\begin{aligned}
I_{\rm GB}&={1\over16\pi}\int_{\cal M}\!\sqrt{-g}\,(R^{\mu\nu\rho\sigma}R_{\mu\nu\rho\sigma}-4R^{\mu\nu}R_{\mu\nu}+R^2)\,d^Dx\cr
&={1\over16\pi}\int_{\cal M}\!\sqrt{-g}\,R^{\mu\nu\rho\sigma}P_{\mu\nu\rho\sigma}\,d^Dx\qquad\hbox{with}\quad  P_{\alpha\beta\gamma\delta}=R_{\alpha\beta\gamma\delta}-2R_{\alpha[\gamma}g_{\delta]\beta}+2R_{\beta[\gamma}g_{\delta]\alpha}+Rg_{\alpha[\gamma}g_{\delta]\beta}\cr
&=
{1\over64\pi}\int_{\cal M}\!\sqrt{-g}\,\delta^{\alpha_1\alpha_2\alpha_3\alpha_4}_{\beta_1\beta_2\beta_3\beta_4}R^{\beta_1\beta_2}_{\ \ \alpha_1\alpha_2}R^{\beta_3\beta_4}_{\ \ \alpha_3\alpha_4}d^Dx\label{GBaction}
\end{aligned}
\end{equation}
where indices are raised by means of the inverse metric $g^{\mu\nu}$, brackets denote antisymmetrization, and $\delta^{\alpha_1\alpha_2\alpha_3\alpha_4}_{\beta_1\beta_2\beta_3\beta_4}$ is the generalized Kronecker symbol, that is, the determinant of the $4\times4$ matrix built from the ordinary Kronecker symbols, with the first row being  $\delta^{\alpha_1}_{\beta_1}$, $\delta^{\alpha_1}_{\beta_2}$ etc. Its variation is obtained using the technology outlined in footnote 3 and is \cite{Myers,GravanisWillison,Davis,DeruelleMadore,DKO}
\begin{equation}
\delta I_{\rm GB}={1\over16\pi}\int_{\cal M}\!\sqrt{-g}\,H_{\mu\nu}\, \delta g^{\mu\nu}\,d^Dx+{\epsilon\over16\pi}\int_{\cal\partial M}\!\left[\sqrt{|h|}B_{ij}\,\delta h^{ij}-\delta\left(\sqrt{|h|}Q_{\rm GB}\right)\right]d^dx\,
\label{GBactionVariation}
\end{equation}
where
\begin{equation}
\begin{aligned}
H^\mu_\nu&=2\left(R^{\mu\alpha\beta\gamma}R_{\nu\alpha\beta\gamma}-2R^{\alpha\beta}R^\mu_{\ \alpha\nu\beta}-2R^{\mu\alpha}R_{\nu\beta}+RR^\mu_\nu\right)-{1\over2}\delta^\mu_\nu\left(R^{\alpha\beta\gamma\delta}R_{\alpha\beta\gamma\delta}-4R^{\alpha\beta}R_{\alpha\beta}+R^2\right)\cr
&=2R^{\mu\beta\gamma\delta}P_{\nu\beta\gamma\delta}-{1\over2}\delta^\mu_\nu\, R^{\alpha\beta\gamma\delta}P_{\alpha\beta\gamma\delta}=-{1\over8}\delta^{\mu\,\alpha_1\alpha_2\alpha_3\alpha_4}_{\nu\,\beta_1\beta_2\beta_3\beta_4}R^{\beta_1\beta_2}_{\ \ \alpha_1\alpha_2}R^{\beta_3\beta_4}_{\ \ \alpha_3\alpha_4} \,, \cr
\end{aligned}
\end{equation}
and where
\begin{equation}
\begin{aligned}
Q_{\rm GB}&=4\delta_{i_1i_2i_3}^{j_1j_2j_3}K^{i_1}_{j_1}\left({1\over2}R^{i_2i_3}_{{\rm b}\,j_2j_3}-{\epsilon\over3}K^{i_2}_{j_2}K^{i_3}_{j_3}\right)=4(J-2G^{\rm b}_{ij}K^{ij})\cr
B^j_i&=2\delta_{i\,i_1i_2i_3}^{j\,j_1j_2j_3}K^{i_1}_{j_1}\left({1\over2}R^{i_2i_3}_{{\rm b}\,j_2j_3}-{\epsilon\over3}K^{i_2}_{j_2}K^{i_3}_{j_3}\right)=2\left(3J^j_i-J\delta^j_i-2P^j_{{\rm b}\,kil}K^{kl}\right)\cr
\hbox{with}\quad \epsilon J_{ij}&=-{2\over3}K_{il}K^{lp}K_{pj}+{2\over3}KK_{il}K^l_j+{1\over3}K_{ij}\left(K^{lp}K_{lp}-K^2\right)\,.\label{GBMboundaryTensor}
\end{aligned}
\end{equation}
$G^{\rm b}_{ij}$ and $P^{\rm b}_{ijkl}$ are the Einstein and $P$ tensors built with the boundary-induced metric $h_{ij}$, and $K_{ij}$ is the extrinsic curvature.

Note that in Eq. (\ref{GBactionVariation}) we omitted the divergence of a four-dimensional vector density in the boundary term $\partial_l(\sqrt{|h|}W^l)$, which is irrelevant since it was evaluated on the closed boundary $\partial{\cal M}$ and studied in Ref. \cite{Padmanabhan}. (Note too that $H^\mu_\nu$ vanishes identically in dimension less than five, as first seen by Bach \cite{Bach} and as its Lovelock expression in terms of the rank-five generalized Kronecker symbol makes obvious.)

As in Einstein gravity, appropriate boundary terms must be added to the Gauss-Bonnet action  in order for the variation of the full action to vanish on shell when Dirichlet boundary conditions ($\delta h^{ij}=0$ on $\partial\cal M$) are imposed. The Gauss-Bonnet-Myers action \cite{Myers}, which generalizes the Einstein Gibbons-Hawking-York one, is
\begin{equation}I_{\rm GBM}=I_{\rm GB}+I_{\rm M}\quad\hbox{with}\quad I_{\rm M}={\epsilon\over16\pi}\int_{\cal\partial M}\!\sqrt{|h|}Q_{\rm GB}d^dx
\,,\label{GBMaction}
\end{equation}
with $Q_{\rm GB}$ given in Eq. (\ref{GBMboundaryTensor}).
It yields a Dirichlet variational principle since (that is, when $H_{\mu\nu}=0$),
\begin{equation}
\delta I_{\rm GBM}\vert_{\rm on shell}={\epsilon\over16\pi}\int_{\cal\partial M}\!\sqrt{|h|}\,B_{ij}\,\delta h^{ij}d^dx\
\label{EGHYactionVariation}
\end{equation}
which vanishes for $\delta h^{ij}=0$.
\\

$\bullet$ {\it In the language of forms}
\medskip

To rewrite the previous expressions in the vielbein language we start again with the Chern-Weil theorem, this time in four dimensions (see Refs. \cite{ChernWeil,MerinoOlea}):
\begin{equation}
\varepsilon_{ABCD}(\Omega^{AB}\wedge\Omega^{CD}-\bar{\Omega}^{AB}\wedge
\bar{\Omega}^{CD})=d\left[  2\varepsilon_{ABCD}\bar{\theta}^{AB}\wedge\left(
\bar{\Omega}^{CD}+{\frac{1}{2}}\bar{D}\bar{\theta}^{CD}+{\frac{1}{3}}\eta
_{EF}\bar{\theta}^{[CE]}\wedge\bar{\theta}^{[FD]}\right)  \right]
\,,\label{ChernWeilTheorem4}%
\end{equation}
where $\bar{\theta}^{AB}=\omega^{AB}-\bar{\omega}^{AB}$,
\begin{align}
\bar{\Omega}^{AB} &  =\mathrm{d}\bar{\omega}^{\left[  AB\right]  }+\eta
_{CD}\bar{\omega}^{\left[  AC\right]  }\wedge\bar{\omega}^{\left[  DB\right]
}\,,\nonumber\\
\bar{D}\bar{\theta}^{\left[  AB\right]  } &  =\mathrm{d}\bar{\theta}^{\left[
AB\right]  }+\eta_{CD}\bar{\omega}^{\left[  AC\right]  }\wedge\bar{\theta
}^{\left[  DB\right]  }+\eta_{CD}\bar{\omega}^{\left[  BC\right]  }\wedge
\bar{\theta}^{\left[  AD\right]  }\,,\label{R_Dtheta_Lorentz}%
\end{align}
and where the expression inside the exterior derivative on the rhs of Eq. (\ref{ChernWeilTheorem4}) is the transgression form.
Antisymmetrization is not needed in these equations if $\bar{\omega}^{AB}$ is antisymmetric.\footnote{In particular, Eq. (\ref{R_Dtheta_Lorentz}) states that if a spin connection $\bar{\omega}^{AB}$ is not antisymmetric, then the curvature $\bar{\Omega}^{AB}$ and covariant derivative $\bar{D}\bar{\theta}^{\left[AB\right]}$ must be contructed only with its antisymmetric part. As shown in Ref. \cite{MerinoOlea}, this ensures that the curvature comes from a well-defined Lorentz gauge connection and  satisfies the Bianchi identities. We also notice that $\bar{\Omega}^{AB}$ is antisymmetric by construction and that the antisymmetrization bracket has been omitted in the derivative term of Eq. (\ref{ChernWeilTheorem4}) because this task is performed by the Levi-Civita tensor which contracts the indices $\left(  CD\right)$.\\} 

The theorem (or identity) (\ref{ChernWeilTheorem4}) again suggests introducing as a \textit{Gauss-Bonnet-Chern-Weil} (GBCW) action its dimensional continuation,  $I_{\rm GBCW}$, defined as (we limit ourselves to five dimensions for notational simplicity)
\begin{equation}
I_{\mathrm{GBCW}}={\frac{1}{64\pi}}\int_{\mathcal{M}_{5}}\!\varepsilon_{ABCDE}\,\Omega^{AB}\wedge\Omega^{CD}\wedge e^{E}-{\frac{1}{32\pi}}\mathrm{d}\int_{\mathcal{M}_{5}}\!\varepsilon_{ABCDE}\,\bar{\theta}^{AB}\wedge\left(  \bar{\Omega}^{CD}+{\frac{1}{2}}\bar{D}\bar{\theta}^{CD}+{\frac{1}{3}}\eta_{EF}\bar{\theta}^{[CE]}\wedge\bar{\theta}^{[FD]}\right)\wedge e^{E}\,,\label{ChernWeilAction5}
\end{equation}
where the first term gives back the Gauss-Bonnet action (\ref{GBaction}),
as it can be easily checked using the properties given in Eq. (\ref{id_translation}) but adapted to a five-dimensional manifold $\mathcal{M}_{5}\,$.
Again, the possible background bulk term $\int\varepsilon_{ABCDE}\,\bar\Omega^{AB}\,\wedge\,\bar\Omega^{CD}\,\wedge e^E$ must be omitted in order to not spoil the field equations.

Now, when the second connection $\bar{\omega}$ describes a product manifold
the term containing the derivative $\bar{D}$ drops out of Eq. (\ref{ChernWeilAction5}
), while the curvature $\bar{\Omega}$ coincides with $\Omega$ in the boundary.
Then, as shown by Myers in Ref. \cite{Myers}, the Gauss-Bonnet-Myers action
(\ref{GBMaction}) can be written in the language of differential forms as%

\begin{equation}
I_{\mathrm{GBM}}={\frac{1}{64\pi}}\int_{\mathcal{M}_{5}}\varepsilon
_{ABCDE}\Omega^{AB}\wedge\Omega^{CD}\wedge e^{E}-{\frac{1}{32\pi}}\mathrm{d}\int_{\mathcal{M}_{5}}\varepsilon_{ABCDE}\,\bar{\theta}^{AB}%
\wedge\left(  \Omega_{\mathrm{b}}^{CD}+{\frac{1}{3}}\eta_{FG}{\bar{\theta}%
}^{CF}{\bar{\theta}}^{GD}\right)  \wedge e^{E}\,,\label{GBMactionForm}%
\end{equation}
where $\Omega_{\mathrm{b}}^{CD}$ is the curvature of the boundary and the
antisymmetrization brackets are not needed because 
the connection $\bar{\omega}$ associated with a product metric is antisymmetric by construction.\\ 

We remark that the translation of Eq. (\ref{GBMactionForm}) to the tensorial version (\ref{GBMaction}) can be made only if the covariance is reduced to the boundary.\footnote{Using the properties given in Sec. III, but adapted to a five-dimensional manifold $\mathcal{M}_{5}\,$, one can directly check that in Gaussian coordinates the boundary term can be written as $-{\frac{1}{32\pi}}\mathrm{d}\int_{\mathcal{M}_{5}}\varepsilon_{ABCDE}\,\bar{\theta}^{AB}\wedge\left(\Omega_{\mathrm{b}}^{CD}+{\frac{1}{3}}\eta_{FG}{\bar{\theta}}^{CF}{\bar{\theta}}^{GD}\right)  \wedge e^{E}={\frac{\epsilon}{16\pi}}\int_{\partial\mathcal{M}_{5}}\varepsilon_{abcd}\,K^{a}\wedge\left(\Omega_{\mathrm{b}}^{bc}-{\frac{\epsilon}{3}}K^{b}K^{c}\right)  \wedge e^{d}$ and the translation of the last expression to tensorial language gives Eq. (\ref{GBMaction}).\\} 
However, as we will see in the next section, the use of the hybrid connection (\ref{hybrid_lorentz}) allows to perform a fully covariant translation of the GBCW action to tensorial language. This will provide not only a way to generalize the Katz procedure to the Einstein-Gauss-Bonnet case, but also the way to write the Myers boundary term in a fully covariant coordinate basis.

\subsection{A Gauss-Bonnet Katz action}

Let us consider the hybrid connection (\ref{hybrid_lorentz}) and its associated curvature two-form $\tilde{\Omega}^{AB}=\mathrm{d}\tilde{\omega}^{AB}+\eta_{CD}\tilde{\omega}^{[AC]}\wedge\tilde{\omega}^{[DB]}$. The relations (see Ref. \cite{MerinoOlea} for further details)
\begin{align}
\tilde{\Omega}^{AB}  & =\Omega^{AB}-\tilde{D}\tilde{\theta
}^{\left[  AB\right]  }-\eta_{CD}\tilde{\theta}^{\left[  AC\right]  }%
\wedge\tilde{\theta}^{\left[  DB\right]  }\,, \nonumber\\
\tilde{D}\tilde{\theta}^{\left[  AB\right]  }  & =D\tilde{\theta}^{\left[
AB\right]  }-2\eta_{CD}\tilde{\theta}^{\left[  AC\right]  }\wedge\tilde{\theta
}^{\left[  DB\right]  }\label{ChernWeilTheorem4_prop}%
\end{align}
can be used to show that the  Chern-Weil action (\ref{ChernWeilAction5}) for the hybrid connection can be expressed as\footnote{We recall that in our conventions, Gauss' theorem for a four-form $Q$ reads $\int_{\mathcal{M}_{5}}dQ=-\epsilon\int_{\partial\mathcal{M}_{5}} Q\,$.\\}
\begin{equation}
\tilde{I}_{\mathrm{GBCW}}=-{\frac{1}{64\pi}}\int_{{\cal M}_5}\!\varepsilon_{ABCDE}\,\Omega^{AB}\wedge\Omega^{CD}+{\frac{\epsilon}{32\pi}}\int_{\partial{\cal M}_5}\!\varepsilon_{ABCDE}\,\tilde{\theta}%
^{AB}\wedge\left(  \Omega^{CD}-{\frac{1}{2}}D\tilde{\theta}^{\left[
CD\right]  }+\frac{1}{3}\eta_{EF}\tilde{\theta}^{\left[  CE\right]  }%
\wedge\tilde{\theta}^{\left[  FD\right]  }\right)  
.\label{ChernWeilAction5bis}%
\end{equation}
The we express $\tilde{\Omega}$ and $\tilde{D}$ in terms of
$\Omega$ and $D$ is that this allows us to translate the transgression form to
tensorial language using $\Omega^{AB}=e_{\mu}^{A}e_{\nu}^{B}\Omega^{\mu\nu}$
and $D\tilde{\theta}^{\left[  AB\right]  }=e_{\mu}^{A}e_{\nu}^{B}\nabla
\tilde{\theta}^{\left[  \mu\nu\right]  }$, where
\begin{align}
\Omega^{\mu\nu}  & =g^{\nu\alpha}\Omega_{\ \alpha}^{\mu}\ ,\ \ \ \ \ \ \Omega
_{\ \alpha}^{\mu}=d\omega_{\ \alpha}^{\mu}+\omega_{\ \sigma}^{\mu}\wedge
\omega_{\ \alpha}^{\sigma}=\frac{1}{2}R_{\ \alpha\beta\gamma}^{\mu}dx^{\beta
}\wedge dx^{\gamma}\,,\nonumber\\
\nabla\tilde{\theta}^{\left[  \mu\nu\right]  }  & =d\tilde{\theta}^{\left[
\mu\nu\right]  }+\omega_{\ \alpha}^{\mu}\wedge\tilde{\theta}^{\left[
\alpha\nu\right]  }+\omega_{\ \alpha}^{\nu}\wedge\tilde{\theta}^{\left[
\mu\alpha\right]  }.\label{R_Dtheta_tensors}%
\end{align}

Building on these results, we are led to propose the Gauss-Bonnet Katz action as the translation of Eq. (\ref{ChernWeilAction5bis}) in a coordinate basis, that is,
\begin{equation}
\begin{aligned}
I_{\mathrm{GBK}} \equiv \tilde{I}_{\mathrm{GBCW}}={\frac{1}{64\pi}}\int_{\mathcal{M}_{5}}\!
E_{\mu\nu\rho\sigma\lambda}\Omega^{\mu\nu}\wedge\Omega^{\rho\sigma
} \wedge\mathrm{d}x^{\lambda}&
 +{\frac{\epsilon}{32\pi}}\int_{\partial\mathcal{M}_{5}} E_{\mu\nu\rho\sigma\alpha
}\tilde{\theta}^{\mu\nu}\wedge\left(  \Omega^{\rho\sigma}-{\frac{1}{2}}%
\nabla\tilde{\theta}^{\left[  \rho\sigma\right]  }+\frac{1}{3}g_{\gamma
\lambda}\tilde{\theta}^{\left[  \rho\gamma\right]  }\tilde{\theta}^{\left[
\lambda\sigma\right]  }\right)  \wedge\mathrm{d}x^{\alpha} \label{GBKActionForm}\\
\hbox{with}\quad\tilde{\theta}^{\mu\nu}  & =\Delta_{\alpha}^{\mu\nu
}\mathrm{d}x^{\alpha}\quad\hbox{where}\quad\Delta_{\alpha}^{\mu\nu}%
=g^{\nu\delta}\Delta_{\alpha\delta}^{\mu}\,. 
\end{aligned} 
\end{equation}
The Gauss-Bonnet Katz vector is then obtained through the relation
\begin{equation}
-d\left[  2E_{\mu\nu\rho\sigma\alpha}\tilde{\theta}^{\mu\nu}\wedge\left(
\Omega^{\rho\sigma}-{\frac{1}{2}}\nabla\tilde{\theta}^{\left[  \rho
\sigma\right]  }+\frac{1}{3}g_{\gamma\lambda}\tilde{\theta}^{\left[
\rho\gamma\right]  }\tilde{\theta}^{\left[  \lambda\sigma\right]  }\right)
\wedge\mathrm{d}x^{\alpha}\right]  =\partial_{\mu}\left(  \sqrt{-g}k_{\rm GB }^{\mu}\right)  d^{5}x\,,
 \end{equation}
which yields~:
\begin{equation}
k_{\rm GB }^{\mu}=-\delta_{\mu_{1}\mu_{2}\mu_{3}\mu_{4}}^{\mu
\nu_{2}\nu_{3}\nu_{4}}\Delta_{\nu_{2}}^{\mu_{1}\mu_{2}}\left( R_{\nu
_{3}\nu_{4}}^{\mu_{3}\mu_{4}}-\nabla_{\nu_{3}}\Delta_{\nu_{4}}^{\left[
\mu_{3}\mu_{4}\right]  }+\frac{2}{3}g_{\gamma\lambda}\Delta_{\nu_{3}}^{\left[
\mu_{3}\gamma\right]  }\Delta_{\nu_{4}}^{\left[  \lambda\mu_{4}\right]
}\right)\,. \label{GBKvector}
\end{equation}

Therefore the Gauss-Bonnet Katz (\ref{GBKActionForm}) action we propose, which generalizes the Einstein Katz action (\ref{EKaction})-(\ref{KboundaryTerm}),
reads, in tensorial language,
\begin{equation}I_{\rm GBK}=I_{\rm GB}+I^{\rm K}_{\rm GB}\quad\hbox{with}\quad I^{\rm K}_{\rm GB}={1\over64\pi}\int_{{\cal M}_5}\!\partial_\mu\left(\sqrt{-g}\,k_{\rm GB}^\mu\right)d^5x \,,\label{GBKaction}
\end{equation}
where $I_{\rm GB}$ and $k_{\rm GB}^\mu$ are defined in Eqs. (\ref{GBaction}) and (\ref{GBKvector}), respectively.
Using Gaussian coordinates the Gauss-Bonnet Katz boundary term also reads\footnote{This result is obtained after expanding the generalized Kronecker
delta with the identity%
\[
\delta_{\mu_{1}\mu_{2}\mu_{3}\mu_{4}}^{\nu_{1}\nu_{2}\nu_{3}\nu_{4}}%
=\delta_{\mu_{1}}^{\nu_{1}}\delta_{\mu_{2}\mu_{3}\mu_{4}}^{\nu_{2}\nu_{3}%
\nu_{4}}-\delta_{\mu_{2}}^{\nu_{1}}\delta_{\mu_{1}\mu_{3}\mu_{4}}^{\nu_{2}%
\nu_{3}\nu_{4}}+\delta_{\mu_{3}}^{\nu_{1}}\delta_{\mu_{1}\mu_{2}\mu_{4}}%
^{\nu_{2}\nu_{3}\nu_{4}}-\delta_{\mu_{4}}^{\nu_{1}}\delta_{\mu_{1}\mu_{2}%
\mu_{3}}^{\nu_{2}\nu_{3}\nu_{4}}\,,
\]
using the Gauss-Codazzi equations $R_{kl}^{ij}=R_{\mathrm{b}\,kl}%
^{ij}-\epsilon\left(  K_{k}^{i}K_{l}^{j}-K_{l}^{i}K_{k}^{j}\right)
\,,$\ $R_{\ kl}^{wi}=-\epsilon\left(  \mathring{\nabla}_{k}K_{l}^{i}%
-\mathring{\nabla}_{l}K_{k}^{i}\right)  $ where $\mathring{\nabla}$ is the
covariant derivative with respect to the connection $\mathring{\Gamma}%
_{jk}^{i}$ of the boundary. Also, we use the following properties:
\[
\nabla_{i_{2}}\Delta_{i_{3}}^{j_{2}j_{3}}=\mathring{\nabla}_{i_{2}}%
\Delta_{i_{3}}^{j_{2}j_{3}}+K_{i_{2}}^{j_{2}}\Delta_{i_{3}}^{wj_{3}}+K_{i_{2}%
}^{j_{3}}\Delta_{i_{3}}^{j_{2}w}+\epsilon K_{i_{2}i_{3}}\Delta_{w}^{j_{2}%
j_{3}}\,,\ \ \ \nabla_{i_{2}}\Delta_{i_{3}}^{\left[  wj_{3}\right]
}=\mathring{\nabla}_{i_{2}}\Delta_{i_{3}}^{\left[  wj_{3}\right]  }-\epsilon
K_{i_{2}l}\Delta_{i_{3}}^{\left[  lj_{3}\right]  }\,,
\]%
\[
\Gamma_{ij}^{w}=-\epsilon K_{ij}\,,\ \ \Gamma_{wj}^{i}=K_{j}^{i}%
\,\,,\ \ \Gamma_{jk}^{i}=\mathring{\Gamma}_{jk}^{i}\,,\ \ \ \Delta_{i}%
^{jl}=\mathring{\Delta}_{i}^{jl}\,,\ \ \ \Delta_{k}^{\left[  wl\right]
}=-\epsilon\left(  K_{k}^{l}+Z_{k}^{l}\right)  \,.
\]
}
\begin{equation}
I_{\mathrm{GB}}^{\mathrm{K}}={\frac{1}{16\pi}}\int_{\partial\mathcal{M}_{5}%
}\sqrt{|h|}\,k_{\mathrm{GB}}^{w}\,d^{4}x\,,\quad\text{with \ \ }k_{\left(
2\right)  }^{w}=\epsilon Q_{\mathrm{GB}}+\epsilon E_{\mathrm{GB}%
}\,,\label{GBKboundaryTerm}%
\end{equation}
where $Q_{\mathrm{GB}}$ is given in Eq. (\ref{GBMboundaryTensor}) and
\begin{align}
E_{\mathrm{GB}} &  =4\delta_{i_{1}i_{2}i_{3}}^{j_{1}j_{2}j_{3}}K_{j_{1}%
}^{i_{1}}\left[  -\frac{1}{2}\mathring{\nabla}_{j_{2}}\mathring{\Delta}%
_{j_{3}}^{i_{2}i_{3}}+\epsilon K_{j_{2}}^{i_{2}}Z_{j_{3}}^{i_{3}}%
-\frac{\epsilon}{3}\left(  K_{j_{2}}^{i_{2}}Z_{j_{3}}^{i_{3}}+Z_{j_{2}}%
^{i_{2}}\left(  K_{j_{3}}^{i_{3}}+Z_{j_{3}}^{i_{3}}\right)  \right)  +\frac
{1}{3}h_{l_{2}l_{3}}\mathring{\Delta}_{j_{2}}^{\left[  i_{2}l_{2}\right]
}\mathring{\Delta}_{j_{3}}^{\left[  l_{3}i_{3}\right]  }\right]  \nonumber\\
&  +4\delta_{i_{1}i_{2}i_{3}}^{j_{1}j_{2}j_{3}}Z_{j_{1}}^{i_{1}}\left[
\frac{1}{2}R_{\mathrm{b}\,j_{2}j_{3}}^{i_{2}i_{3}}-\frac{1}{2}\mathring
{\nabla}_{j_{2}}\mathring{\Delta}_{j_{3}}^{i_{2}i_{3}}+\epsilon K_{j_{2}%
}^{i_{2}}Z_{j_{3}}^{i_{3}}-\frac{\epsilon}{3}\left(  K_{j_{2}}^{i_{2}%
}+Z_{j_{2}}^{i_{2}}\right)  \left(  K_{j_{3}}^{i_{3}}+Z_{j_{3}}^{i_{3}%
}\right)  +\frac{1}{3}h_{l_{2}l_{3}}\mathring{\Delta}_{j_{2}}^{\left[
i_{2}l_{2}\right]  }\mathring{\Delta}_{j_{3}}^{\left[  l_{3}i_{3}\right]
}\right]  \nonumber\\
&  -2\delta_{i_{1}i_{2}i_{3}}^{j_{1}j_{2}j_{3}}\mathring{\Delta}_{j_{1}%
}^{i_{1}i_{2}}\left[  \mathring{\nabla}_{j_{2}}\left(  Z_{j_{3}}^{i_{3}%
}-K_{j_{3}}^{i_{3}}\right)  +K_{j_{2}l_{3}}\mathring{\Delta}_{j_{3}}^{\left[
l_{3}i_{3}\right]  }-\frac{2}{3}h_{l_{2}l_{3}}\left(  K_{j_{2}}^{l_{2}%
}+Z_{j_{2}}^{l_{2}}\right)  \mathring{\Delta}_{j_{3}}^{\left[  l_{3}%
i_{3}\right]  }\right] \,, \label{BGK_Extra_term}%
\end{align}
where $Z_{j}^{i}=-\frac{1}{2}\left(  h^{ik}+\bar{h}^{ik}\right)  \bar{K}_{jk}\,$, $\mathring{\nabla}$ is the covariant derivative with respect to the connection $\mathring{\Gamma}_{jk}^{i}$ of the boundary, $\mathring{\Delta}_{j}^{il}=h^{lk}\mathring{\Delta}_{jk}^{i}$, and $\mathring{\Delta}_{jk}^{i}=\mathring{\Gamma}_{jk}^{i}-\overline{\mathring{\Gamma}}_{jk}^{i}\,$.\\

Similarly to the results found in Sec. \ref{sectionIII}, the Gauss-Bonnet Katz action (\ref{GBKActionForm}) and (\ref{GBKboundaryTerm}) reproduces the Gauss-Bonnet-Myers action (\ref{GBMaction}) if the background is replaced by a product manifold. In that case the extra term $E_{\mathrm{GB}}$ vanishes and thus $I_{\rm GBK}=I_{\rm GBM}$, because as the boundaries coincide and $\bar{K}_{ij}=0$ one gets
$\Delta_{j}^{ik}=0$ and $Z_{j}^{i}=0$. 
This means that the action (\ref{GBKActionForm}) also represents the way to write the Gauss-Bonnet-Myers action (\ref{GBMaction}) in a fully covariant coordinate basis by only replacing the background by a product manifold.
To show this one must use the relations $\mathrm{d}%
x^{\alpha}\wedge\mathrm{d}x^{\beta}\wedge\mathrm{d}x^{\gamma}\wedge
\mathrm{d}x^{\delta}\wedge\mathrm{d}x^{\lambda}=-\varepsilon^{\alpha
\beta\gamma\delta\lambda}d^{5}x$,  
$\varepsilon_{\mu\nu\rho\sigma\lambda
}\varepsilon^{\alpha\beta\gamma\delta\lambda}=-\delta_{\mu\nu\rho\sigma
}^{\alpha\beta\gamma\delta}$, $\varepsilon^{ijkl}\varepsilon_{spql}%
=-\epsilon\,\delta_{spq}^{ijk}$,  
and ${\tilde{\theta}}^{\mu\nu}=g^{\nu\gamma}\left(  \omega_{\ \gamma}^{\mu}-\bar{\omega}_{\ \gamma}^{\mu}\right)$, with
$\omega_{\ \gamma}^{\mu}$ and $\bar{\omega}_{\ \gamma}^{\mu}$ being defined in Eq. (\ref{other_hyb_def}). 
We note for further reference that at the boundary $dw=0$ and ${\tilde{\theta}}^{ij}=0\,$, because in the product manifold case the
boundaries coincide so that $\Gamma_{lk}^{i}=\bar{\Gamma}_{lk}^{i}\,$. 
Then the normal index $w$ can only be in ${\tilde{\theta}}^{\mu\nu}\,$, while the
quadratic term inside the parentheses gives $g_{\gamma\lambda}{\tilde{\theta}%
}^{\left[  j\gamma\right]  }{\tilde{\theta}}^{\left[  \lambda k\right]
}=g_{ww}{\tilde{\theta}}^{\left[  jw\right]  }{\tilde{\theta}}^{\left[
wk\right]  }=-\epsilon K_{p}^{j}K_{q}^{k}dx^{p}dx^{q}\,$, where we have used
$g_{ww}=\epsilon$  and  ${\tilde{\theta}}^{\left[  iw\right]  }=-{\tilde{\theta
}}^{\left[  wi\right]  }=\epsilon K_{l}^{i}dx^{l}$ which can be easily
obtained using $\Gamma_{ij}^{w}=-\epsilon K_{ij}\,$, $\Gamma_{jw}^{i}=K_{j}^{i}$,
and $\bar{\Gamma}_{ij}^{w}=\bar{\Gamma}_{jw}^{i}=0$ because $\bar{K}%
_{ij}=\frac{1}{2}\partial_{w}\bar{h}_{ij}=0$ for a product manifold. Thus, all of the information about the product manifold disappears, 
giving us the known tensorial result (\ref{GBMaction}) for $D=5$.\\

The proposals for the Gauss-Bonnet Katz action (\ref{GBKActionForm}) and (\ref{GBKaction}), the Katz vector (\ref{GBKvector}), and the boundary term (\ref{GBKboundaryTerm}) form the core of this paper.

\subsection{A Dirichlet variational principle}

Here we show that the Gauss-Bonnet-Katz action $I_{\mathrm{GBK}}$ proposed in the previous section solves the Dirichlet problem. 
There are \textit{a priori} various ways to show this, as one can work using its tensorial form (\ref{GBKaction}) or its expression in 
differential-form language in a coordinate basis (\ref{GBKActionForm}) or using Eq. (\ref{ChernWeilAction5bis}).
The proof we will give here is based on the dimensional continuation of the variation of the transgression form, which can be obtained using homotopic techniques in terms of an interpolating connection.

As shown in the Appendix, the Chern-Weil theorem for the hybrid connection can be written as
$\varepsilon_{ABCD}\Omega^{AB}\Omega^{CD}-\varepsilon_{ABCD}\tilde{\Omega
}^{AB}\tilde{\Omega}^{CD}=d\mathcal{T}^{\left(  3\right)  }\left(
\omega,\tilde{\omega}\right)  \,$, where the transgression form is given by (we omit the $\wedge$ symbols)
\begin{equation}
\mathcal{T}^{\left(  3\right)  }\left(  \omega,\tilde{\omega}\right)
=2\int_{0}^{1}dt\varepsilon_{ABCD}\tilde{\theta}^{AB}\Omega_{\left(  t\right)
}^{CD}\,,\label{Dir_proof_forms_1}%
\end{equation}
where $\Omega_{\left(  t\right)  }^{AB}=d\omega_{\left(  t\right)  }^{AB}
+\omega_{\left(  t\right)  C}^{A}\omega_{\left(  t\right)  }^{CB}\,\ $ and
$\omega_{\left(  t\right)  }^{AB}=\tilde{\omega}^{\left[  AB\right]  }%
+t\tilde{\theta}^{\left[  AB\right]  }$ (with $\tilde{\theta}^{AB}=\omega
^{AB}-\tilde{\omega}^{AB}$) is a connection interpolating between
$\tilde{\omega}^{\left[  AB\right]  }\,$ and $\omega^{AB}$. It is also shown in the Appendix
that its variation is then given by
\begin{equation}
\delta\mathcal{T}^{\left(  3\right)  }\left(  \omega,\tilde{\omega}\right)
=2\varepsilon_{ABCD}\left(  \Omega^{AB}\delta\omega^{CD}-\tilde{\Omega}%
^{AB}\delta\tilde{\omega}^{CD}\right)  -2d\int_{0}^{1}dt\varepsilon
_{ABCD}\tilde{\theta}^{AB}\delta\omega_{\left(  t\right)  }^{CD}%
\,.\label{Dir_proof_forms_2}%
\end{equation}

Then, the Gauss-Bonnet action and the corresponding boundary term $I_{\mathrm{GB}%
}^{\mathrm{K}}$ we have proposed in Eqs. (\ref{GBKActionForm}) and
(\ref{GBKaction}) can be written in the vielbein language as
\[
I_{\mathrm{GB}}=\frac{1}{64\pi}\int_{\mathcal{M}_{5}}\varepsilon_{ABCDE}%
\Omega^{AB}\Omega^{CD}e^{E}\,,\ \ \ \ \ I_{\mathrm{GB}}^{\mathrm{K}}=\frac
{1}{64\pi}\int_{\mathcal{M}_{5}}d\beta_{\left(  2\right)  }\,,
\]
where $\beta_{\left(  2\right)  }$ is the dimensional continuation of the
transgression,
\begin{equation}
\beta_{\left(  2\right)  }\left(  \omega,\tilde{\omega}\right)  =-2\int
_{0}^{1}dt\varepsilon_{ABCDE}\tilde{\theta}^{AB}\Omega_{\left(  t\right)
}^{CD}e^{E}\,.\label{Dir_proof_forms_3}%
\end{equation}
To show that the variation of $I_{\mathrm{GBK}}=I_{\mathrm{GB}}+I_{\mathrm{GB}}^{\mathrm{K}}$ vanishes for Dirichlet
conditions we see that $\delta\beta_{\left(  2\right)  }$ must be the
dimensional continuation of $\delta\mathcal{T}^{\left(  3\right)  }\left(
\omega,\tilde{\omega}\right)  $ given in Eq. (\ref{Dir_proof_forms_2}) plus an
additional term containing the variation of the vielbein arising in Eq. (\ref{Dir_proof_forms_3}) from the dimensional continuation procedure. Thus we
have $\delta I_{\mathrm{GB}}^{\mathrm{K}}=\frac{1}{64\pi}\int_{\mathcal{M}%
_{5}}d\left(  \delta\beta_{\left(  2\right)  }\right)  $, where
\begin{align}
\delta\beta_{\left(  2\right)  } &  =-2\varepsilon_{ABCDE}\Omega^{AB}%
\delta\omega^{CD}e^{E}+2\varepsilon_{ABCDE}\tilde{\Omega}^{AB}\delta
\tilde{\omega}^{CD}e^{E}\nonumber\\
&  +2d\int_{0}^{1}dt\varepsilon_{ABCDE}\theta^{AB}\delta\omega_{t}^{CD}%
e^{E}-2\int_{0}^{1}dt\varepsilon_{ABCDE}\theta^{AB}\Omega_{\left(  t\right)
}^{CD}\delta e^{E}\,.\label{Dir_proof_forms_4}%
\end{align}
Under the exterior derivative, the first term in Eq. (\ref{Dir_proof_forms_4})
will cancel the variation of the Gauss-Bonnet action,%
\begin{equation}
\delta I_{\mathrm{GB}}=\frac{1}{32\pi}\int_{\mathcal{M}_{5}}d\left(
\varepsilon_{ABCDE}\Omega^{AB}\delta\omega^{CD}e^{E}\right)
\,.\label{Dir_proof_forms_5}%
\end{equation}
The fourth term vanishes for Dirichlet conditions on the boundary ($\left.
\delta e^{E}\right\vert _{\partial\mathcal{M}_{5}}=0$), and the third one
trivially vanishes because $d^{2}=0$. Thus, we have
\[
\delta I_{\mathrm{GB}}^{\mathrm{K}}=-\frac{1}{32\pi}\int_{\mathcal{M}_{5}%
}d\left(  \varepsilon_{ABCDE}\left(  \Omega^{AB}\delta\omega^{CD}%
-\tilde{\Omega}^{AB}\delta\tilde{\omega}^{CD}\right)  e^{E}\right)  -\frac
{1}{32\pi}\int_{\mathcal{M}_{5}}d\left(  \int_{0}^{1}dt\varepsilon
_{ABCDE}\theta^{AB}\Omega_{\left(  t\right)  }^{CD}\delta e^{E}\right)\,.
\]

Now we only have to prove that the exterior derivative of the second term
vanishes for Dirichlet conditions. Indeed, the antisymmetric part of the
hybrid connection can be written as $\tilde{\omega}^{\left[
AB\right]  }\,=e_{\alpha}^{\left[  A\right.  }\bar{\nabla}e^{\left.  B\right]
\alpha}$, where $\bar{\nabla}=dx^{\mu}\bar{\nabla}_{\mu}$ is the covariant
derivative with respect to the background connection $\bar{\Gamma}$. Taking
$\delta\bar{\Gamma}=0$, we obtain%
\begin{equation}
d\left[  2\varepsilon_{ABCDE}\tilde{\Omega}^{AB}\delta\tilde{\omega}^{CD}%
e^{E}\right]  =d\left[  2\varepsilon_{ABCDE}\tilde{\Omega}^{AB}\bar{\nabla
}e^{D\alpha}e^{E}\delta e_{\alpha}^{C}\right]  +d\left[  2\varepsilon
_{ABCDE}\tilde{\Omega}^{AB}e_{\alpha}^{C}\bar{\nabla}\delta e^{D\alpha}%
e^{E}\right]\,.  \label{di_2}%
\end{equation}
Now, using
\[
d\left(  2\varepsilon_{ABCDE}\tilde{\Omega}^{AB}e_{\alpha}^{C}\delta
e^{D\alpha}e^{E}\right)  =2\varepsilon_{ABCDE}\left[  \bar{\nabla}\left(
\tilde{\Omega}^{AB}e_{\alpha}^{C}e^{E}\right)  \delta e^{D\alpha}%
+\tilde{\Omega}^{AB}e_{\alpha}^{C}\bar{\nabla}\left(  \delta e^{D\alpha
}\right)  e^{E}\right]
\]
and $d^{2}=0$, we get
\begin{equation}
d\left[  2\varepsilon_{ABCDE}\tilde{\Omega}^{AB}\delta\tilde{\omega}^{CD}%
e^{E}\right]  =d\left[  2\varepsilon_{ABCDE}\left(  \tilde{\Omega}^{AB}%
\bar{\nabla}e^{D\alpha}e^{E}\delta e_{\alpha}^{C}-\bar{\nabla}\left(
\tilde{\Omega}^{AB}e_{\alpha}^{C}e^{E}\right)  \delta e^{D\alpha}\right)
\right]\,.
\label{di_3}%
\end{equation}
Then, the variation of $I_{\mathrm{GB}}^{\mathrm{K}}$ is%
\begin{align}
\delta I_{\mathrm{GB}}^{\mathrm{K}} &  =-\frac{1}{32\pi}\int_{\mathcal{M}_{5}%
}d\left(  \varepsilon_{ABCDE}\Omega^{AB}\delta\omega^{CD}e^{E}\right)
\nonumber\\
&  +\frac{1}{32\pi}\int_{\mathcal{M}_{5}}d\left[  \varepsilon_{ABCDE}\left(
\tilde{\Omega}^{AB}\bar{\nabla}e^{D\alpha}e^{E}\delta e_{\alpha}^{C}%
-\bar{\nabla}\left(  \tilde{\Omega}^{AB}e_{\alpha}^{C}e^{E}\right)  \delta
e^{D\alpha}-\int_{0}^{1}dt\theta^{AB}R_{t}^{CD}\delta e^{E}\right)  \right]\,,
\label{di_4}%
\end{align}
so that the variation of $I_{\mathrm{GBK}}=I_{\mathrm{GB}%
}+I_{\mathrm{GB}}^{\mathrm{K}}$ is given by%
\begin{equation}
\delta I_{\mathrm{GBK}}=2\int_{\mathcal{M}_{5}}d\left[  \varepsilon
_{ABCDE}\left(  \tilde{\Omega}^{AB}\bar{\nabla}e^{D\alpha}e^{E}\delta
e_{\alpha}^{C}-\bar{\nabla}\left(  \tilde{\Omega}^{AB}e_{\alpha}^{C}%
e^{E}\right)  \delta e^{D\alpha}-\int_{0}^{1}dt\theta^{AB}R_{t}^{CD}\delta
e^{E}\right)  \right]  \,,\label{di_6}%
\end{equation}
which has a suitable form to apply Dirichlet boundary conditions on the
vielbein, $\left.  \delta e^{A}\right\vert _{\partial\mathcal{M}_{5}}=0\,$. Q.E.D.

\subsection{Einstein-Gauss-Bonnet-Katz action and the Boulware-Deser black hole}

The static, spherically symmetric black hole Boulware-Deser metric – which solves the  Einstein-Gauss-Bonnet equations of motion $G_{\mu\nu}+\alpha H_{\mu\nu}=0$ in five dimensions – is
\begin{align*}
ds^{2}  & =-f\left(  r\right)  \,c^{2}dt^{2}+\frac{1}{f\left(  r\right)
}dr^{2}+r^{2}d\Omega_{\left(  D-2\right)  }^{2}\,,\\
f\left(  r\right)    & =1+\frac{r^{2}}{2\tilde{\alpha}}-\frac{r^{2}}%
{2\tilde{\alpha}}\sqrt{1-\frac{4\tilde{\alpha}}{\ell^{2}}+\frac{4\mu}{r^{D-1}%
}}\,,
\end{align*}
where $\ell^{2}=-\frac{\left(  D-1\right)  \left(  D-2\right)  }{2\Lambda}\,$,
$\tilde{\alpha}=\left(  D-3\right)  \left(  D-4\right)\alpha $, and $\mu$ is the
mass parameter. Its mass has been obtained by various methods, in particular \textit{à la} Katz in Ref. \cite{DKO}. In Eq. (3.9) of that reference the
following Gauss-Bonnet-Katz vector was proposed
\begin{equation}
k_{\mathrm{DKO}}^{\mu}=4R_{\ \ \ \sigma}^{\mu\nu\rho}\Delta_{\nu\rho}^{\sigma
}-8\left(  R^{\mu\nu}\Delta_{\nu\rho}^{\rho}-R^{\nu\rho}\Delta_{\nu\rho}^{\mu
}\right)  +2R\left(  g^{\mu\nu}\Delta_{\nu\rho}^{\rho}-g^{\nu\rho}\Delta
_{\nu\rho}^{\mu}\right)  \,.\label{DKO_vector}%
\end{equation}
This led to the right mass (when taking the background to be asymptotically AdS, i.e., when $\mu$ is taken to be zero in $f(r)$), although it was noted that it did not yield a
well-posed Dirichlet problem. To make a comparison with our construction, we can write Eq. (\ref{GBKvector}) as\footnote{This can be obtained by expanding the Kronecker delta in Eq. (\ref{GBKvector}) with the following identity
\[
\delta_{\mu_{1}\mu_{2}\mu_{3}\mu_{4}}^{\nu_{1}\nu_{2}\nu_{3}\nu_{4}}%
=\delta_{\mu_{1}\mu_{2}}^{\nu_{1}\nu_{2}}\delta_{\mu_{3}\mu_{4}}^{\nu_{3}%
\nu_{4}}-\delta_{\mu_{1}\mu_{3}}^{\nu_{1}\nu_{2}}\delta_{\mu_{2}\mu_{4}}%
^{\nu_{3}\nu_{4}}+\delta_{\mu_{1}\mu_{4}}^{\nu_{1}\nu_{2}}\delta_{\mu_{2}%
\mu_{3}}^{\nu_{3}\nu_{4}}+\delta_{\mu_{2}\mu_{3}}^{\nu_{1}\nu_{2}}\delta
_{\mu_{1}\mu_{4}}^{\nu_{3}\nu_{4}}-\delta_{\mu_{2}\mu_{4}}^{\nu_{1}\nu_{2}%
}\delta_{\mu_{1}\mu_{3}}^{\nu_{3}\nu_{4}}+\delta_{\mu_{3}\mu_{4}}^{\nu_{1}%
\nu_{2}}\delta_{\mu_{1}\mu_{2}}^{\nu_{3}\nu_{4}}\,,
\]
where $\delta_{\mu_{1}\mu_{2}}^{\nu_{1}\nu_{2}}=\delta_{\mu_{1}}^{\nu_{1}%
}\delta_{\mu_{2}}^{\nu_{2}}-\delta_{\mu_{2}}^{\nu_{1}}\delta_{\mu_{1}}%
^{\nu_{2}}$.}
\begin{align}
k_{\mathrm{GB}}^{\mu}  & =4R_{\ \ \ \sigma}^{\mu\nu\rho}\Delta_{\nu\rho
}^{\sigma}-4\left(  R^{\mu\nu}\Delta_{\nu\rho}^{\rho}-R^{\nu\rho}\Delta
_{\nu\rho}^{\mu}+R_{\ \sigma}^{\nu}g^{\mu\rho}\Delta_{\nu\rho}^{\sigma
}-R_{\ \sigma}^{\mu}g^{\nu\rho}\Delta_{\nu\rho}^{\sigma}\right)  +2R\left(
g^{\mu\nu}\Delta_{\nu\rho}^{\rho}-g^{\nu\rho}\Delta_{\nu\rho}^{\mu}\right)
\nonumber\\
& +4\Delta_{\nu}^{\left[  \mu\nu\right]  }\nabla_{\rho}\Delta_{\sigma
}^{\left[  \rho\sigma\right]  }+4\Delta_{\nu}^{\left[  \rho\sigma\right]
}\nabla_{\rho}\Delta_{\sigma}^{\left[  \mu\nu\right]  }+8\Delta_{\sigma
}^{\left[  \rho\sigma\right]  }\nabla_{\left[  \nu\right.  }\Delta_{\left.
\rho\right]  }^{\left[  \mu\nu\right]  }+8\Delta_{\rho}^{\left[  \mu
\nu\right]  }\nabla_{\left[  \sigma\right.  }\Delta_{\left.  \nu\right]
}^{\left[  \rho\sigma\right]  }\nonumber\\
& -\frac{8}{3}g_{\gamma\lambda}\left(  \Delta_{\left[  \rho\right.  }^{\left[
\rho\gamma\right]  }\Delta_{\left.  \sigma\right]  }^{\left[  \lambda
\sigma\right]  }\Delta_{\nu}^{\left[  \mu\nu\right]  }-2\Delta_{\rho}^{\left[
\mu\nu\right]  }\Delta_{\left[  \nu\right.  }^{\left[  \rho\gamma\right]
}\Delta_{\left.  \sigma\right]  }^{\left[  \lambda\sigma\right]  }%
+2\Delta_{\rho}^{\left[  \rho\sigma\right]  }\Delta_{\left[  \sigma\right.
}^{\left[  \mu\gamma\right]  }\Delta_{\left.  \nu\right]  }^{\left[
\lambda\nu\right]  }+\Delta_{\nu}^{\left[  \rho\sigma\right]  }\Delta_{\rho
}^{\left[  \mu\gamma\right]  }\Delta_{\sigma}^{\left[  \lambda\nu\right]
}\right)\,.
\label{kGB_mu_expanded}%
\end{align}
It is an easy (Mathematica) exercise to redo the calculation with the Gauss-Bonnet-Katz vector proposed in this paper and to verify that it also
gives the right mass. Interestingly, only the terms of the first line in Eq. (\ref{kGB_mu_expanded}) (which is different from $k_{\mathrm{DKO}}^{\mu}$)
contribute to the mass, while the precise combination of the derivative and cubic terms ensure that the Dirichlet problem is well defined.

\section{Conclusions}

In the context of background-substraction methods, we have proposed  a boundary term for Einstein-Gauss-Bonnet gravity, which both solves the Dirichlet problem and ensures the finiteness of the conserved charges in this theory. The Chern-Weil theorem for the hybrid connection gives rise to a mathematical structure at the boundary, which naturally accommodates the Christoffel connection associated to the background spacetime. 
We have also shown that the use of such an object is essential in the construction of Katz-like terms in general relativity and Einstein-Gauss-Bonnet theory.
 
In a way, what we have explored here is the use of an enhanced symmetry (Lorentz in the tangent space) to fix the ambiguity of the possible boundary terms that sets a well-posed action principle. This leads to  a generalized Katz vector in EGB gravity such that the correct mass for the Boulware-Deser black hole is obtained.
Our results, based on the use of the Chern-Weil theorem show that the vector that simultaneously allows to obtain, e.g., the right mass of the Boulware-Deser black hole and defines the Dirichlet problem is unique.

The construction presented in this work allows to deal with a general background which can be chosen so as to yield finite global charges without invoking extra regularization procedures. At the same time, it includes the GHM boundary term as a special case when the background is taken to be a product manifold. 

Our analysis naturally suggests that a generalization of the Katz procedure for a generic Lovelock gravity can be performed and the results we have for that problem will be reported soon \cite{MerinoOlea}.

\section*{Acknowledgements}
The authors are grateful to Andr\'es Anabal\'on, F\'elix-Louis Juli\'e and Olivera Miskovic for useful comments. N.M. is supported by a Becas-Chile postdoctoral grant.
The work of R.O. is funded in part by FONDECYT Grant No. 1170765, UNAB Grant DI-1336-16/R, and CONICYT Grant DPI 20140115.

\appendix

\section{Chern-Weil theorem in $D=4$ and variation of the transgression form}

Let us define the connection%
\[
\omega_{\left(  t\right)  }^{AB}=\tilde{\omega}^{\left[  AB\right]  }%
+t\tilde{\theta}^{\left[  AB\right]  }\,,\ \ \ \tilde{\theta}^{AB}=\omega
^{AB}-\tilde{\omega}^{AB}\,,
\]
which interpolates between $\tilde{\omega}^{\left[AB\right]  }\,$ and $\omega^{AB}$. Using $\tilde{D}\tilde{\theta}^{\left[  AB\right]  }
=D\tilde{\theta}^{\left[  AB\right]  }-2\eta_{FG}\tilde{\theta}^{\left[
AF\right]  }\tilde{\theta}^{\left[  GB\right]  }\,$the interpolating field strength
 $\Omega_{\left(  t\right)  }^{AB}=d\omega_{\left(  t\right)  }%
^{AB}+\omega_{\left(  t\right)  C}^{A}\omega_{\left(  t\right)  }^{CB}$ can be
written as%
\begin{align*}
\Omega_{\left(  t\right)  }^{AB} &  =\tilde{\Omega}^{AB}+t\tilde{D}%
\tilde{\theta}^{\left[  AB\right]  }+t^{2}\eta_{FG}\tilde{\theta}^{\left[
AF\right]  }\tilde{\theta}^{\left[  GB\right]  }\\
&  =\Omega^{AB}+\left(  t-1\right)  D\tilde{\theta}^{\left[  AB\right]
}+\left(  t-1\right)  ^{2}\eta_{FG}\tilde{\theta}^{\left[  AF\right]  }%
\tilde{\theta}^{\left[  GB\right]  }\,,
\end{align*}
which gives $\Omega_{\left(  0\right)  }^{AB}=\tilde{\Omega}^{AB}$\ and
$\Omega_{\left(  1\right)  }^{AB}=\Omega^{AB}$. With these definitions, the four-dimensional Chern-Weil theorem given in Eq. (\ref{ChernWeilTheorem4}) can
be written as $\varepsilon_{ABCD}\Omega^{AB}\Omega^{CD}-\varepsilon
_{ABCD}\tilde{\Omega}^{AB}\tilde{\Omega}^{CD}=d\mathcal{T}^{\left(  3\right)
}\left(  \omega,\tilde{\omega}\right)  \,$, so that the transgression form is
given by
\begin{equation}
\mathcal{T}^{\left(  3\right)  }\left(  \omega,\tilde{\omega}\right)
=2\int_{0}^{1}dt\varepsilon_{ABCD}\tilde{\theta}^{AB}\Omega_{\left(  t\right)
}^{CD}\,.\label{CW_2d_2}%
\end{equation}

To calculate the variation $\delta\mathcal{T}^{\left(  3\right)  }$ we need the properties
\begin{align}
\delta\Omega_{\left(  t\right)  }^{AB} &  =D^{\left(  t\right)  }\left(
\delta\omega_{\left(  t\right)  }^{AB}\right)  \,,\ \ \ \delta\omega_{\left(
t\right)  }^{AB}=\delta\tilde{\omega}^{\left[  AB\right]  }+t\delta
\tilde{\theta}^{\left[  AB\right]  }\,,\ \ \ \tilde{\theta}^{\left[
AB\right]  }=\frac{d}{dt}\left(  \omega_{\left(  t\right)  }^{AB}\right)
\,,\nonumber\\
\delta\tilde{\theta}^{\left[  AB\right]  } &  =\frac{d}{dt}\left(
\delta\omega_{\left(  t\right)  }^{AB}\right)  \,,\ \ \ D^{\left(  t\right)
}\tilde{\theta}^{\left[  AB\right]  }=\frac{d}{dt}\left(  \Omega_{\left(
t\right)  }^{AB}\right)  \,,\label{tran_properties}%
\end{align}
which can be easily obtained from the previous definitions. Thus, we have
\begin{align}
\delta\mathcal{T}^{\left(  3\right)  }\left(  \omega,\tilde{\omega}\right)
&  =2\int_{0}^{1}dt\varepsilon_{ABCD}\left(  \delta\tilde{\theta}^{AB}%
\Omega_{\left(  t\right)  }^{CD}+\tilde{\theta}^{AB}D^{\left(  t\right)
}\left(  \delta\omega_{\left(  t\right)  }^{CD}\right)  \right)  \nonumber\\
&  =2\int_{0}^{1}dt\varepsilon_{ABCD}\left(  \delta\tilde{\theta}^{AB}%
\Omega_{\left(  t\right)  }^{CD}+\frac{d}{dt}\left(  \Omega_{\left(  t\right)
}^{AB}\right)  \delta\omega_{\left(  t\right)  }^{CD}\right)  -2d\int_{0}%
^{1}dt\left[  \varepsilon_{ABCD}\tilde{\theta}^{AB}\delta\omega_{\left(
t\right)  }^{CD}\right]  \,,\label{step_1}%
\end{align}
where we have used%
\[
\varepsilon_{ABCD}\tilde{\theta}^{AB}D^{\left(  t\right)  }\left(
\delta\omega_{\left(  t\right)  }^{CD}\right)  =\varepsilon_{ABCD}D^{\left(
t\right)  }\tilde{\theta}^{AB}\delta\omega_{\left(  t\right)  }^{CD}-d\left[
\varepsilon_{ABCD}\tilde{\theta}^{AB}\delta\omega_{\left(  t\right)  }%
^{CD}\right]
\]
and some of the properties given in Eq. (\ref{tran_properties}). The terms in the first integral in Eq. (\ref{step_1}) give
\[
\varepsilon_{ABCD}\left(  \delta\tilde{\theta}^{AB}\Omega_{\left(  t\right)
}^{CD}+\frac{d}{dt}\left(  \Omega_{\left(  t\right)  }^{AB}\right)
\delta\omega_{\left(  t\right)  }^{CD}\right)  =\varepsilon_{ABCD}\frac{d}%
{dt}\left(  t\Omega_{\left(  t\right)  }^{AB}\right)  \delta\tilde{\theta
}^{CD}+\varepsilon_{ABCD}\frac{d}{dt}\left(  \Omega_{\left(  t\right)  }%
^{AB}\right)  \delta\tilde{\omega}^{CD}\,,
\]
and thus we obtain
\begin{equation}
\delta\mathcal{T}^{\left(  3\right)  }\left(  \omega,\tilde{\omega}\right)
=2\varepsilon_{ABCD}\left(  \Omega^{AB}\delta\omega^{CD}-\tilde{\Omega}%
^{AB}\delta\tilde{\omega}^{CD}\right)  -2d\int_{0}^{1}dt\varepsilon
_{ABCD}\tilde{\theta}^{AB}\delta\omega_{\left(  t\right)  }^{CD}%
\,.\label{var_tran_4d}%
\end{equation}

This is the relation used in the main text to show that the Gauss-Bonnet-Katz action indeed yields a Dirichlet variational principle. Note that the use of the interpolating connection $\omega_{\left(  t\right)  }^{AB}$ is
essential to get this result.

\end{document}